\documentclass[useAMS,usenatbib]{mn2e}
 \usepackage[dvips]{graphicx}
\title[Environmental and mass dependencies of $\lambda$]{Environment and mass dependencies of galactic $\lambda$ spin
parameter: cosmological simulations and observed galaxies compared}
\author[B. Cervantes-Sodi, X. Hernandez, Changbom Park and Juhan Kim]{B. Cervantes-Sodi$^{1}$\thanks{E-mail:
bcervant@astroscu.unam.mx}, X. Hernandez$^{1}$\thanks{E-mail: xavier@astroscu.unam.mx}, Changbom Park$^{2}$ and Juhan Kim$^{2}$\\
$^{1}$Instituto de Astronom\'\i a,
Universidad Nacional Aut\'onoma de M\'exico
A. P. 70--264,  M\'exico 04510 D.F., M\'exico \\
$^{2}$Korea Institute for Advanced Study, Dongdaemun-gu, Seoul 130-722, Korea\\
}
\begin{document}

\date{In original form 2007 November 11}

\pagerange{\pageref{863}--\pageref{872}} \pubyear{2008}

\maketitle

\label{firstpage}

\begin{abstract}
We use a sample of galaxies from the Sloan Digital Sky Survey (SDSS) to search for correlations between the $\lambda$ 
spin parameter and the environment and mass of galaxies. In order to calculate the total value of $\lambda$ 
for each observed galaxy, we employed a simple model of the dynamical structure of the galaxies, which allows
a rough estimate of the value of $\lambda$ using only readily obtainable observables from the luminous galaxies. 
Use of a large volume-limited sample (upwards of 11,000) allows reliable inferences of mean values and dispersions 
of $\lambda$ distributions. We find, in agreement with some
N-body cosmological simulations, no significant dependence of $\lambda$ on the environmental density of the galaxies.
For the case of mass, our results show a marked correlation with $\lambda$, in the sense that low-mass
galaxies present both higher mean values of $\lambda$ and associated dispersions, than high-mass galaxies. These results provide interesting 
constrain on the mechanisms of galaxy formation and acquisition of angular momentum, a valuable test for cosmological models.
\end{abstract}


\section{Introduction}

One of the most studied parameters in numerical simulations of
formation and evolution of galaxies is the $\lambda$ spin parameter,
first introduced by Peebles in 1969, where it was used to test the
gravitational instability picture as a theory for the origin of
galaxies. Since then, there have been several studies using
the $\lambda$ parameter as an indispensable tool of analysis, or characterizing its
distribution in cosmological N-body simulations. As the first numerical simulations were done,
some estimates of mean values of $\lambda$ were obtained (Peebles 1971, Efstathiou \&
Jones 1979, Blumenthal et al. 1984, Barnes \& Efstathiou 1987),
as well as probability distributions. In general, the distributions
of this parameter are well fitted by a lognormal function, characterized
by two parameters; $\lambda_{0}$, the most probable value,
and $\sigma_{\lambda}$, which accounts for the spread of the
distribution. In a recent work, Shaw et al. (2006)
showed a compilation of results for estimates of these parameters, arising from numerical
studies performed by several authors, lying in the ranges: $0.03<\lambda_{0}<0.05$
and $0.48<\sigma_{\lambda}<0.66$. Besides numerical simulations, there have been several
analytical attempts (Peebles 1969, Doroshkevich 1970, White 1984, Heavens \& Peacock 1988)
at predicting $\lambda_{0}$ and $\sigma _{\lambda}$ with similar results. More
recently, Syer, Mao \& Mo (1999) confirmed the form of the distribution for $\lambda$
using observational data of 2500 objects, and some of us in a previous work
(Hernandez et al. 2007), using the same 
sample of 11 597 galaxies from the Sloan Digital Sky Survey (SDSS) we treat here, obtained
the distribution of this parameter, with results well fitted by a lognormal function with
parameters $\lambda_{0}=0.04\pm 0.005$ and $\sigma_{\lambda}=0.51 \pm 0.05$, for the first time
derived from a statistical sample of real galaxies. Using the same model, Puech et al. (2007) 
obtained a similar result using a sample of intermediate-redshift galaxies.

The origin of galactic angular momentum is commonly explained as a result of
tidal torques of neighboring protogalaxies on the forming galactic halo. As the protogalaxy breaks away from the
general expansion of the Universe, and since in the general case, it is unlikely
to be spherically symmetric, the forming galaxy is torqued up through coupling with the ambient tidal field. This picture appeared 
with Hoyle in 1949 and was developed at first order
analytically where the growth of the angular momentum is proportional to time, and later explored using
numerical N-body simulations where the complicated processes of gravitational interactions
can be tracked (e.g. Sugerman, Summers \& Kamionkowski 2000). Other
explanations for the origin of the angular momentum have been proposed, like the model
of Vitvitska et al. (2002), in which the haloes obtain their spin through the cumulative
acquisition of angular momentum from satellite accretion, obtaining distributions well modelled
by lognormal functions with parameters similar to the ones obtained with the tidal torque theory.
Variations to the general theory have  been tested by e.g. Maller, Dekel \& Somerville (2002), proposing
mainly two scenarios; in the first, the halo spin is generated by the transfer of orbital angular momentum
from satellites that merge with the main halo, and a second one were linear tidal-torque theory is
applied to shells of infalling matter. The evolution for $\lambda$ at different redshift is completely different for both scenarios, as well as
the dependence with mass, where a trend in the tidal-torque scenario is clear, in the sense that
more massive galaxies tend to present low $\lambda$ values and low dispersions. The study by Barnes \& Efstathiou (1987)
revealed a weak trend towards decreasing $\lambda_{0}$ with increasing mass, confirmed in some N-body
simulations (Cole \& Lacey, 1996, Bett et al. 2007) and not found in others (Maccio et al. 2007).

A direct measure of the real distribution of galactic $\lambda$ distributions, as a function of environment
density and mass, would therefore constrain and inform theories of angular momentum acquisition and galaxy formation.
It is precisely this that we attempt, using a first-order estimate of galactic halo $\lambda$, to derive
distributions of this parameter as functions of mass and environment density from a large SDSS sample, and compare
against results from large cosmological N-body simulations.

The existence of a correlation between galaxy morphology and local density environment (Dressler 1980,
Goto et al. 2003, Park et al. 2007) is a motivation to search for other correlations between local environment and
internal properties of galaxies. Lemson \& Kauffmann (1999), using an N-body simulation, found that
the only quantity which varies as a function of environment was the mass distribution, and Maccio et al.
(2007) confirmed an absence of correlation between $\lambda$ and environment in cosmological simulations, 
although the opposite is sometimes claimed by other studies using similar techniques (e.g. Avila-Reese et al. 2005).
Even the spin alignment of dark matter haloes in different environments, has been studied in some N-body simulations
(Arag\'on-Calvo et al. 2007). In this work, we present a study using a sample of galaxies from the SDSS, in an attempt
to give an observational counterpart to such studies performed using N-body simulations. 

The paper is organized as follows. In Section 2, we present a review of the derivation of the $\lambda$
parameter for infered haloes of any galaxy, spiral or elliptical, developed in Hernandez \& Cervantes-Sodi (2006) (Paper I) and 
Hernandez et al. (2007), henceforth Paper II. In Section 3 we introduce the sample used to perform the study, 
and there we present our general results, including a comparison against a recent high-resolution cosmological 
N-body simulation. Finally, in Section 4 we present a discussion of the results and the final conclusions.

\section{Estimation of $\lambda$ from observable parameters}

The currently accepted picture for galaxy formation is the Lambda cold dark matter ($\Lambda$CDM) model where dark matter overdensities in 
the expanding Universe at high redshift, accrete baryonic material through their gravitational potential, and via
gravitational evolution grow to become galaxies. Their principal
integral characteristics, according to theoretical studies, are their mass and angular momentum. In this section 
we briefly summarize the hypothesis behind the estimates derived in Papers I and II, used to estimate $\lambda$
for the haloes of observed galaxies, from observable parameters. A comparison of our estimate with
results coming from numerical simulations is also included. The following are intended only as first order estimates. 
The errors for individual galaxies are of order 30 per cent (see below), it is only through the use of extensive samples that
meaningful inferences on the distribution of $\lambda$ can be derived.

\subsection{Estimates of halo $\lambda$ parameters from observed galactic properties}

The angular momentum is commonly characterized by the dimensionless angular momentum parameter

\begin{equation}
\label{Lamdef}
\lambda = \frac{L \mid E \mid^{1/2}}{G M^{5/2}}
\end{equation}

where $E$, $M$ and $L$ are the total energy, mass and angular momentum of the configuration, respectively. In 
Paper I we derived a simple estimate of $\lambda$ for disc galaxies in terms of observational parameters, and 
showed some clear correlations between this parameter and structural parameters, as the disc to bulge ratio,
the scale height and the colour, after estimating  $\lambda$ for a sample of real galaxies. Here we recall briefly the main ingredients of the simple model. The model 
consider only two components, a disc for the baryonic component with an exponential surface mass density $\Sigma(r)$:

\begin{equation}
\label{Expprof}
\Sigma(r)=\Sigma_{0} e^{-r/R_{d}},
\end{equation} 

where $r$ is a radial coordinate and $\Sigma_{0}$ and $R_{d}$ are two constants which are allowed to vary from
galaxy to galaxy, and a dark matter halo having an isothermal density profile $\rho(r)$, responsible for establishing
a rigorously flat rotation curve $V_{d}$ throughout the entire galaxy:

\begin{equation}
\label{RhoHalo}
\rho(r)={{1}\over{4 \pi G}}  \left( {{V_{d}}\over{r}} \right)^{2}. 
\end{equation}

We assume that (1) that the specific angular momentum of the disc and halo are equal, e.g. Fall \& Efstathiou (1980), Mo, Mao \& White (1998); 
(2) the total energy is dominated by that of the halo which is a virialized gravitational structure; (3) the disc 
mass is a constant fraction of the halo mass $F=M_{d}/M_{H}$. These assumptions allow us to express $\lambda$ as

\begin{equation}
\label{Lamhalo}
\lambda=\frac{2^{1/2} V_{d}^{2} R_{d}}{G M_{H}}.
\end{equation}

Finally, we introduce a disc Tully-Fisher (TF) relation: $M_{d}=A_{TF} V_{d}^{3.5}$, and taking the Milky Way as a 
representative example, we evaluate $F$ and $A_{TF}$ to obtain

\begin{equation}
\label{LamObs}
\lambda=21.8 \frac{R_{d}/kpc}{(V_{d}/km s^{-1})^{3/2}}.
\end{equation}

In Paper II, we presented a derivation for an equivalent expression
to equation ~\ref{LamObs} for elliptical galaxies, again using a model of two components: a baryonic matter 
distribution with a Hernquist density profile and a dark matter halo, in principle, showing no difference
from that of disc galaxies (White \& Rees 1978; Kashlinsky 1982). Following this hypothesis, the energy
can be obtained from the density profile described by equation ~\ref{RhoHalo} assuming again a virialized
structure, which allow us to calculate the mass at the half light radius, $R_{50}$:

\begin{equation}
M_{dyn}=\frac{(1.65\sigma)^{2}R_{50}}{G},
\end{equation}

from Padmanabhan et al. (2004), where $\sigma$ is the velocity dispersion. For the angular momentum, again we 
suppose that the specific angular momentum of dark matter and baryons are equal, and using dimensional analysis
we expect that it will be proportional to the eccentricity of the system, $e$; and the factor $(GM_{b}a)^{1/2}$,
where $M_{b}$ is the baryonic mass and $a$ is the mayor axis. Introducing a numerical factor to account for the
dissipation of the baryonic component, its dynamical pressure support and the projection effects, we finally obtain

\begin{equation}
\label{LamObsEll}
\lambda= \frac{0.1173 e (a/kpc)^{2/7}}{(\sigma/km s^{-1})^{3/7}}.
\end{equation}

With equations ~\ref{LamObs} and ~\ref{LamObsEll}, we can obtain an estimate of $\lambda$ for the halo of any galaxy using just 
the most basic information available. 

\subsection{Testing the model against detailed galactic simulations}

To check our estimate of $\lambda$ we made use of numerical simulations where this parameter acts as an input 
parameter, or its value can be estimated with high accuracy.

The model by Hernandez, Avila-Reese \& Firmani (2001) includes a very wide range of physics, initial conditions are supplied by
an statistical sampling of a cosmological primordial fluctuation spectrum, giving a mass aggregation history
of gas and dark matter. No TF type of relation is assumed a priori, indeed, this codes
attempt to recover such a relation as a final result of the physics included.
A fixed value of $\lambda$ is assumed for the infalling material which settles into a disc
in centrifugal equilibrium and differential rotation, viscous friction results in radial flows of 
matter and angular momentum. The redistribution of mass affects the rotation curve in a self-consistent way, through
a Poisson equation including the disc self gravity and a dark halo which responds to mass redistributions through
an adiabatic contraction. The star formation is followed in detail, with an energy balance cycle.

\begin{figure}
\includegraphics[width=84mm]{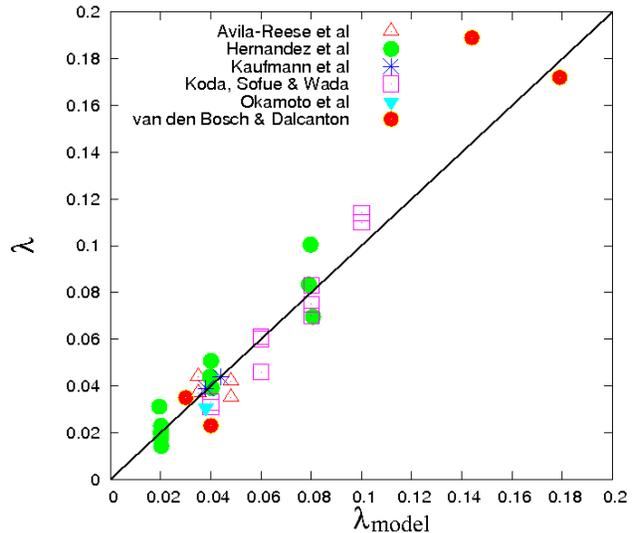}
 \caption{Comparison between the calculated $\lambda$ value using equation ~\ref{LamObs} and the actual value,
using simulated galaxies from different groups.}
  \label{models}
\end{figure}

A simple population synthesis code then traces the luminosities and colours of the various stellar populations at all 
radii and times. The van den Bosch \& Dalcanton (2000) models are qualitatively similar to the previous ones, but vary in numerical 
approaches, resolution, time step-issues and the level of approximation and inclusion of many different physical 
aspects of the complicated problem.

Kaufmann et al. (2007) explore the angular momentum transport, disc morphology and how radial profiles 
depend sensitively on force and mass resolution. They study systematically this effect with controlled N-body/smoothed 
hydrodynamics simulations of disc galaxy formation by cooling a rotating gaseous mass distribution inside equilibrium 
cuspy spherical and triaxial dark matter haloes employing up to $10^{6}$ gas and dark matter particles. 
Their models are calibrated using the Milky Way and M33 models for the comparisons. Varied and detailed physical effects
are introduced self-consistently in their simulations, from the end results of which we use resulting galactic properties
to estimate halo $\lambda$ using equation (\ref{LamObs}), and compare to their input values.

Avila-Reese et al. (2005) employed a $\Lambda CDM$ N-body simulation to study the properties of galaxy-size dark 
matter haloes as a function of global environment, where $\lambda$ was one of the studied properties of the halo, 
in particular its value in different environments such as clusters, voids or field. We then used reported values 
of the final resulting galactic properties, disc scale lengths and rotation curves to estimate the $\lambda$
parameters through equation (\ref{LamObs}), and compare against the input values used in the simulations.

With the aim of studying the origin of the TF relation, Koda, Sofue \& Wada (2000), used a N-body/smoothed particle
hydrodynamic method, including cooling, star formation, and stellar feedback of energy, mass and metals, to trace
the evolution of the galaxies from $z=25$ to $z=0$ in a CDM cosmology. The resulting spiral galaxies, with exponential 
disc profiles and generically flat rotation curves, including disc self-gravity and which reproduce the slope 
and scatter of the TF relation, all these as results of the evolved physics, not input suppositions, were reported.

The work of Okamoto et al. (2005), was performed using hydrodynamic simulations of galaxy formation in a $\Lambda CDM$ 
universe with different models of star formation and feedback, to see their effect on the morphology of disc galaxies.
Also giving as output present day galactic properties from which we estimated halo $\lambda$ through equation (\ref{LamObs}),
and compared against their input model value.

The comparison between the input $\lambda$ values of the galaxies modelled by the different groups ($\lambda_{model}$), 
and our estimate using equation (\ref{LamObs}) is presented in Fig. \ref{models}, where the x-axis value correspond 
to the exact value coming from the detailed study of the system, and the y-axis value is our estimate using 'observable' 
parameters through equation (\ref{LamObs}). 
We can appreciate a very good agreement between the calculated value with our simple model, and the actual value of 
$\lambda$ for simulated galaxies, regardless of the complexity of the models used. While all the above models include fairly 
detailed physics, it is remarkable that our simple dimensional estimate agrees so well, allways to better than 30 per cent, 
in most cases much better. As no systematics
are apparent, we can now use equation (\ref{LamObs}) with a sample of real galaxies, to obtain statistical properties of the
distributions of $\lambda$ in the real universe.

\section{Comparisons of $\lambda$ distributions from the SDSS and cosmological N-body simulations}

\subsection{Observational sample}

The sample of real galaxies employed for the study comes from the SDSS Data Release 5 (Adelman-McCarthy et al. 2007). It is a volume-limited sample having galaxies in the redshifts interval $0.025 < z < 0.055$ and absolute magnitudes $M_{r}-5 log h \leq -18.5$. Since most of the studies concerning spin distributions from simulations presents their 
results at $z=0$, we limited the sample to low redshifts. This sample contains 32 550 galaxies
for which Choi, Park \& Vogeley (2007) have determined the exponential disc scales, absolute magnitudes, velocity
dispersions, de Vacouleurs radii and seeing corrected isophotal ellipticities for each galaxy, assuming a $\Lambda CDM$ universe with 
$\Omega_{M}=0.27$, $\Omega_{\Lambda}=0.73$ and $h=0.71$. To obtain $\lambda$ for each galaxy, we need to discriminate
between elliptical and disc galaxies; in order to do that we used the prescription of Park \& Choi (2005), in 
which early (ellipticals and lenticulars) and late (spirals and irregulars) types are segregated in a $u - r$ colour versus $g - i$ 
colour gradient space and in the concentration index space. In order to apply equation ~\ref{LamObs} to the disc 
galaxies of the sample, we need the rotation velocity, which is inferred from the absolute magnitude using a TF 
relation (Barton et al. 2001), then, to avoid the problem of internal absorption in edge-on galaxies, we employed only spiral galaxies having axis ratios $> 0.6$ and inferred rotation velocities in the range of $80<V_{R}<430$, well within the range of applicability of the TF realtion we are using. After applying these two cuts, and removing a randomly selected fraction of ellipticals, so as to maintain the original early- to late-type fraction, we are left with a total of 11 597 galaxies.

The measurement of $R_{d}$ for our SDSS galaxies comes form the total distribution of light, without any decomposition of the light coming from the bulge or the disc, in this way, systems with prominent bulges, where an important fraction of mass is present in the bulge with low angular momentum, present short $R_{d}$ values, which imply a low $\lambda$ value assigned trough equation ~\ref{LamObs}, accounting implicitly for the low angular momentum of the bulge fraction. We note that of the simulated galaxies in Fig. 1, Avila-Reese et al. (2005) and Koda et al. (2000) include prominent bulge components. It is reassuring of the implicit bulge accounting described above that no systematics are apparent in Fig. 1, regarding either the early-type simulated galaxies, or any other ones. For the elliptical galaxies we apply equation ~\ref{LamObsEll}, as fully described in Paper II, using the corrected isophotal ellipticities to calculate the eccentricity.

\subsection{Simulation \& Subhalo Finding}
To compare our observational sample with a numerical simulation,
we have made a cosmological N-body simulation of a $\Lambda$CDM model
in a cubic box with the side length of $614 h^{-1} {\rm Mpc}$. The model
parameters are $h=0.7$, $\Omega_m=0.27$, $\Omega_b=0.046$, and
$\Omega_\Lambda=0.73$, given by the \textit{Wilkinson Microwave Anisotropy Probe} 1-year data (Spergel et al. 2003).
The simulation gravitationally evolved $1024^3$ CDM particles from redshift
$z=48$ to $z=0$ taking 1880 global time-steps.
The linear density fluctuations of matter at the present epoch is normalized
by  $\sigma_8 = 0.9$, the RMS density fluctuation at the $8h^{-1}$Mpc top-hat
smoothing scale.
The mass of each simulation particle is $M_p = 1.4 \times 10^{10} h^{-1}{\rm M_\odot}$
and the force resolution is about $60 h^{-1}kpc$.

To identify subhalos in the simulation,
we first extracted the dark matter particles located in virialized regions
by applying the standard friend-of-friend (FOF) method
with a linking length equal to one-fifth the mean particle separation. This
length is a characteristic scale for the identification of virialized
structures. To each FOF particle group, we applied the PSB method
(Kim \& Park 2006; Kim, Park \& Choi 2008) to find subhalos.
The method identifies subhalos that are gravitationally self-bound
in terms of the total energy and stable against the external tidal force.
The resulting subhalo population is complete down to
$M_h = 4.3 \times 10^{11} h^{-1} {\rm M_\odot}$. This has been
found from a comparison of the subhalo mass function with that of the base
mass functions of subhalos obtained from other higher resolution simulations
(Kim \& Park 2008, in preparation). The minimum halo mass amounts
to a collective mass of 30 particles and corresponds to an early-type galaxy
with $M_r=-19$ in the subhalo-galaxy correspondence model, which assumes
each subhalo contains one and only one optical galaxy (Kim et al. 2008). Henceforth, the term halo, when applied to our simulations, will refer to subhalo, as defined above.

For comparisons with
our observational results we randomly selected 100 000 halos from the simulation, for which the spin parameter $\lambda$ was calculated numerically. The normalized local
density parameter is obtained following the same method as used for our
SDSS sample (see section 3.4 for more details,
see also Park et al. 2007; Kim et al. 2008).

\begin{figure}
\centering
\begin{tabular}{c}
\includegraphics[width=0.475\textwidth]{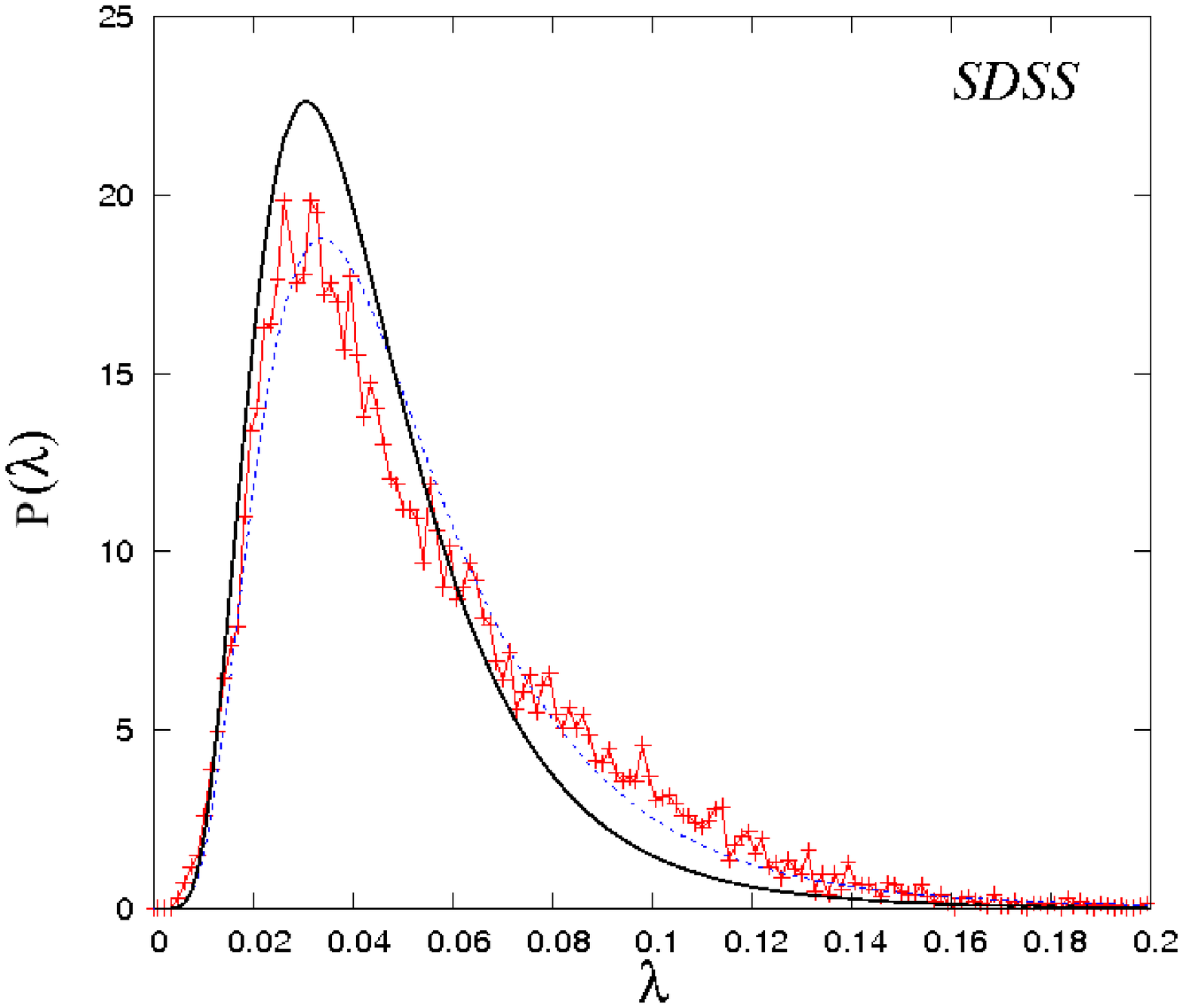}\\
\includegraphics[width=0.475\textwidth]{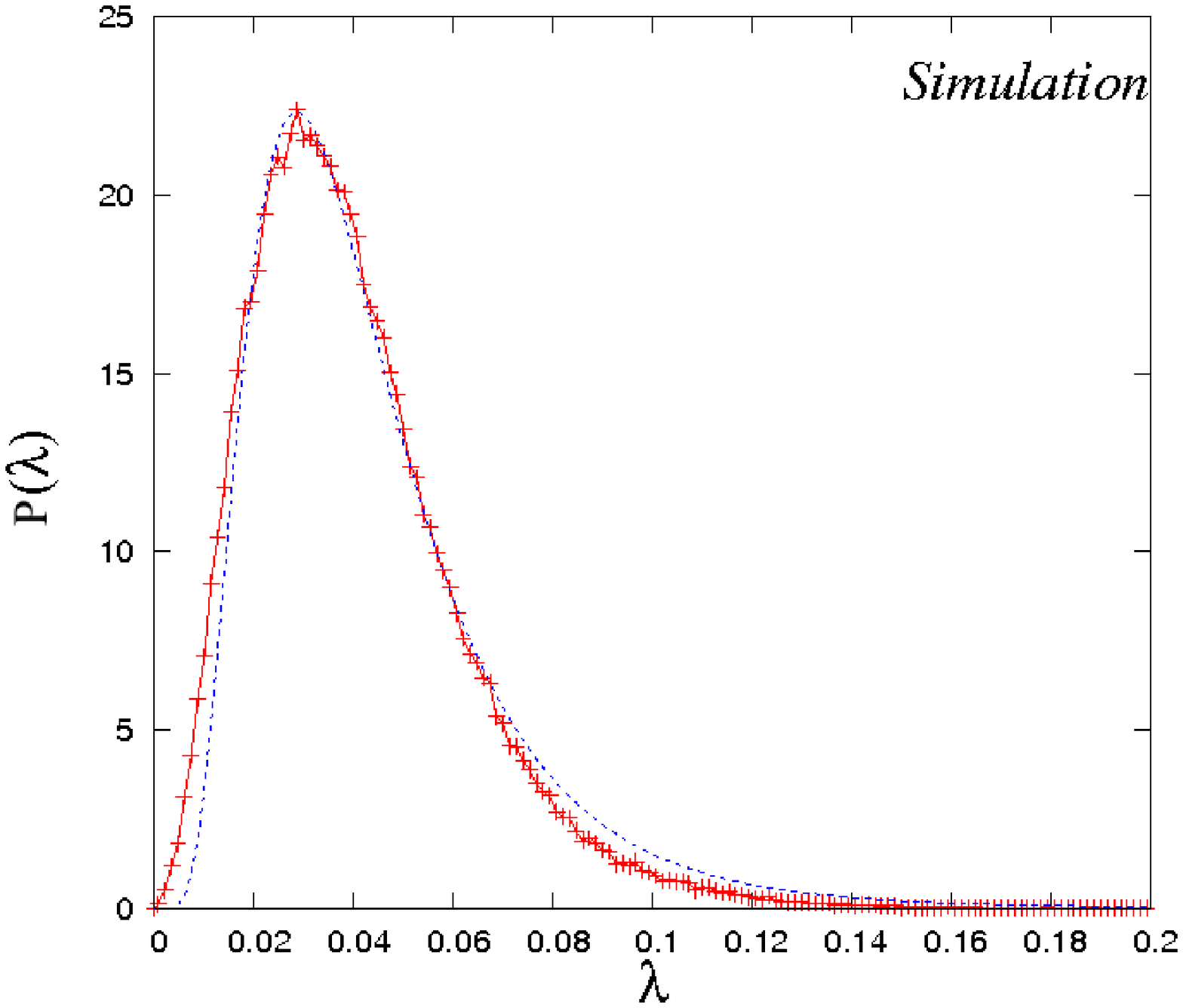}
\end{tabular}
\label{histograms}
\caption[ ]{Distribution of $\lambda$ values, broken curve corresponding to the data binned into 150 intervals and
 dotted curve best lognormal fit to the data. Top: Data from the SDSS sample using 11 597 galaxies, with 
parameters  $\lambda_{0}=0.046$ and $\sigma_{\lambda}=0.535$, solid curve being the maximum likelihood underlying distribution, assuming a $30$ error in our inferred $\lambda$ values, with parameters  $\lambda_{0}=0.0394$ and $\sigma_{\lambda}=0.509$. Bottom: Data from the N-body simulation 
using 100 000 simulated dark matter haloes, with parameters  $\lambda_{0}=0.038$ and $\sigma_{\lambda}=0.525$.}
\end{figure}

\subsection{Integral properties of the $\lambda$ distributions}

In Fig. 2, we present the distribution of the spin parameter for the SDSS and the simulated samples; 
the broken curves show histograms after dividing the data into 150 bins, and the dotted curves give the best direct fit to a 
lognormal distribution of the form

\begin{equation}
\label{Plam}
P(\lambda_{0},\sigma_{\lambda};\lambda) d\lambda=
\frac{1}{\sigma_{\lambda}\sqrt{2\pi}}exp\left[-\frac{ln^{2}(\lambda/\lambda_{0})}
{2\sigma_{\lambda}^{2}} \right] \frac{d\lambda}{\lambda}.
\end{equation}

We can appreciate that the distributions are well fitted by this function, where the corresponding values from the 
SDSS and the simulation are: $\lambda_{0}=0.046$, $\sigma_{\lambda}=0.535$ and $\lambda_{0}=0.038$, $\sigma_{\lambda}=0.525$, 
both results in agreement with previous predictions from N-body simulations (from a recent compilation for $\lambda_{0}$ and $\sigma_{\lambda}$ values, 
see Shaw et al. 2006).

The empirical scaling relations used for our estimation of $\lambda$ show a measure of dispersion. In the case that these dispersions were intrinsic to the data, they would propagate as errors in our $\lambda$ estimates. In order to estimate what is the best underlying distribution, assuming that the distribution of the data is a degraded version of the intrinsic lognormal distribution, in Paper II we employed a full maximum likelihood analysis, which gave us a best fit with parameters  $\lambda_{0}=0.0394$ and $\sigma_{\lambda}=0.509$, showed in Fig. 2, top panel, as a solid line. The maximum likelihood fit and the direct fit over the data from the sample are extreme cases, one assuming that the dispresions in the empirical scalings used are fully intrinsic to the data, and the other that they are strictly measurements errors, the actual $\lambda$ distribution should lie between these two cases, in very good agreement with the distribution from the simulation, we can hence see that the integral distribution of galactic
halo $\lambda$ spin parameters in real galaxies is consistent with what current structure formation theories predict.

That both the $\lambda$ distributions inferred from the SDSS sample, and those coming from all published cosmological
simulations are well fitted by lognormal distributions, is interesting. We note that lognormal distributions often arise as a
consequence of the central limit theorem. Whenever the particular outcome of a variable is the sum of a large number
of independent distributions, we are guaranteed that the final distribution will be a Gaussian.
If the particular outcome of a variable is the product of a large number of
independent distributions, it follows that the logarithm of the variable will be normally distributed. What the many
independent distributions might in this case be, we can only speculate, the repeated merger events leading to the formation of
a galaxy, or perhaps the tidal fields due to surrounding density fields as they sequentially detach from the overall 
expansion of the universe and virialize.

\begin{figure*}
\centering
\begin{tabular}{cc}
\includegraphics[width=0.475\textwidth]{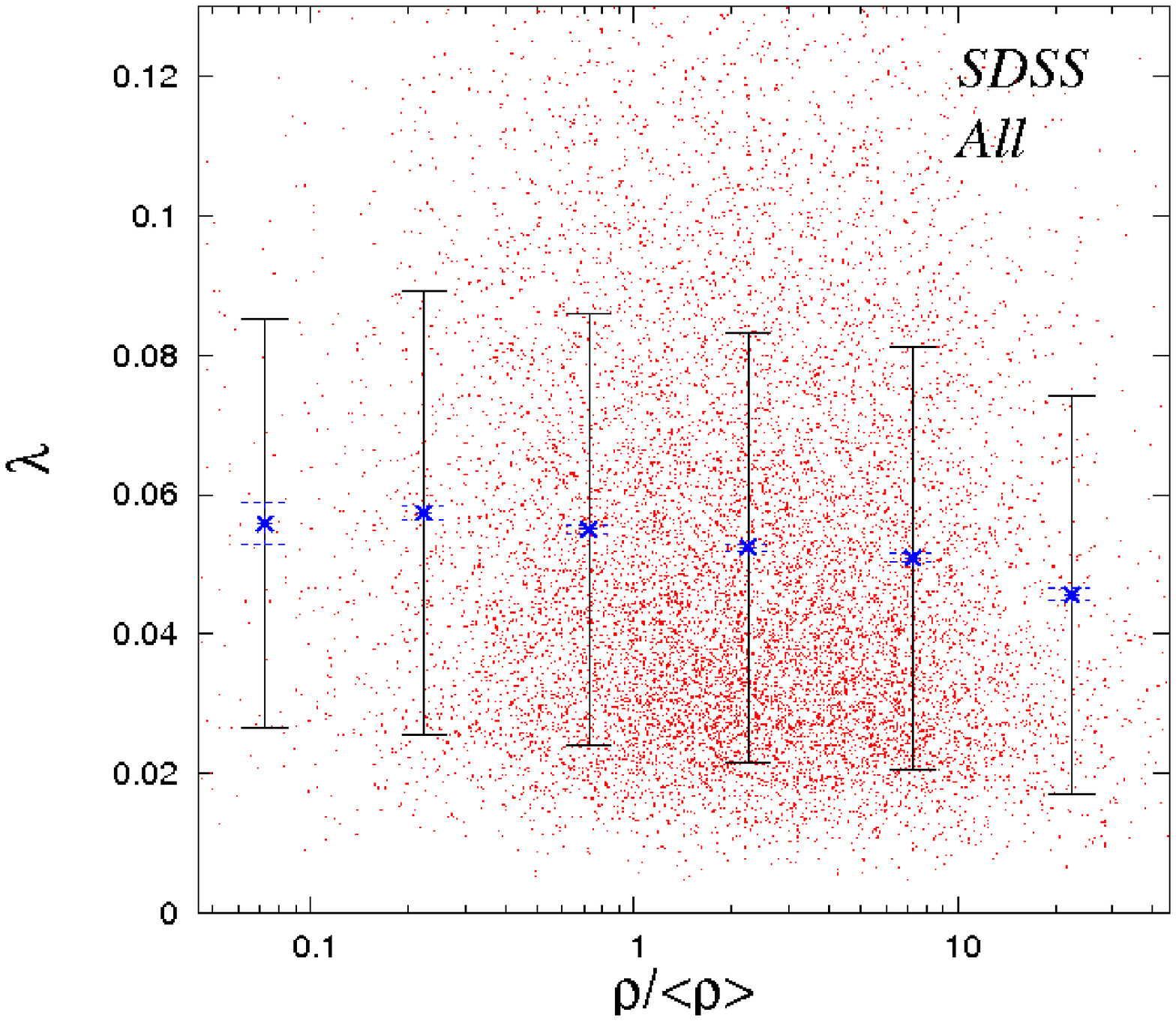} & \includegraphics[width=0.475\textwidth]{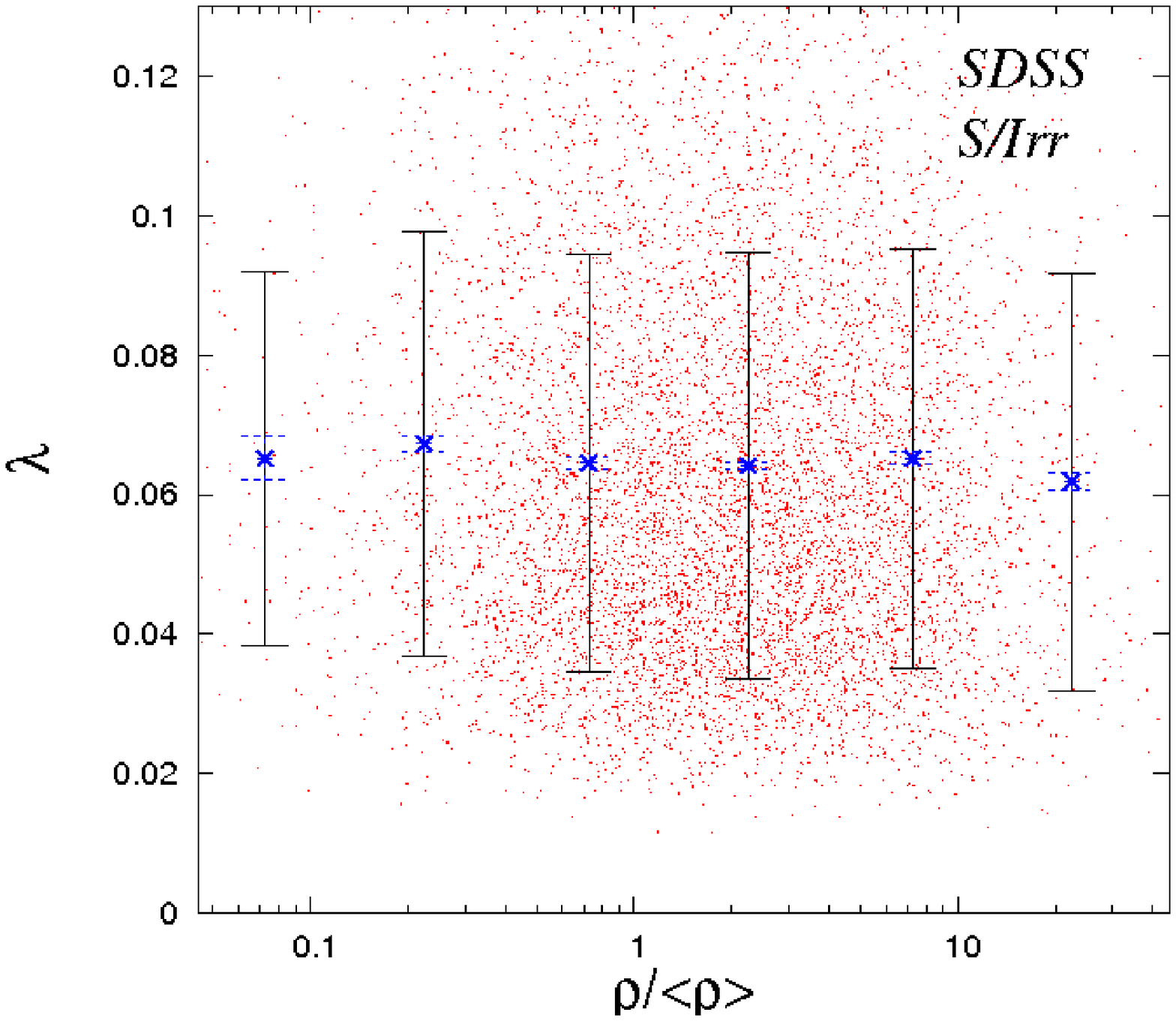} \\
\includegraphics[width=0.475\textwidth]{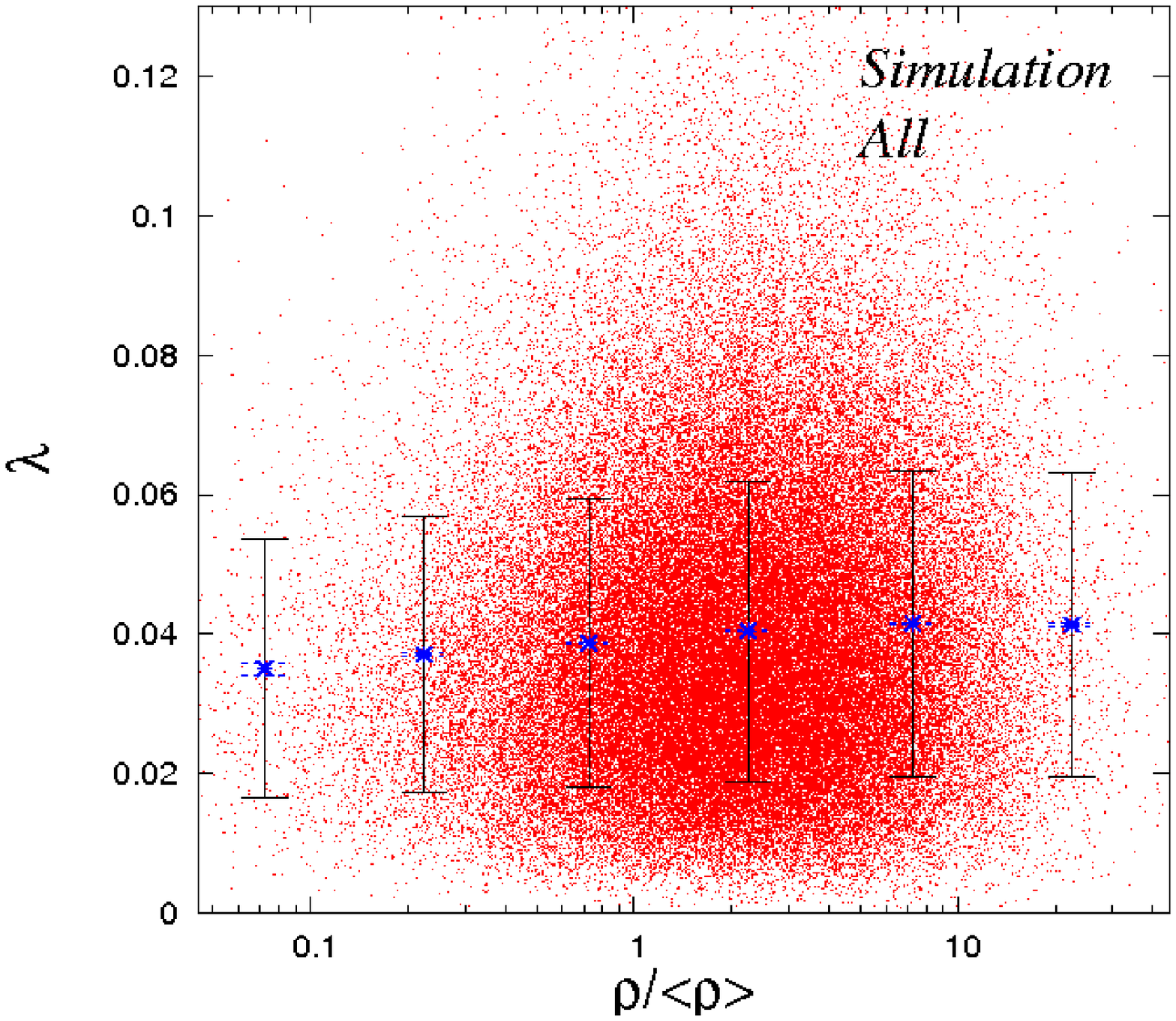} & \includegraphics[width=0.475\textwidth]{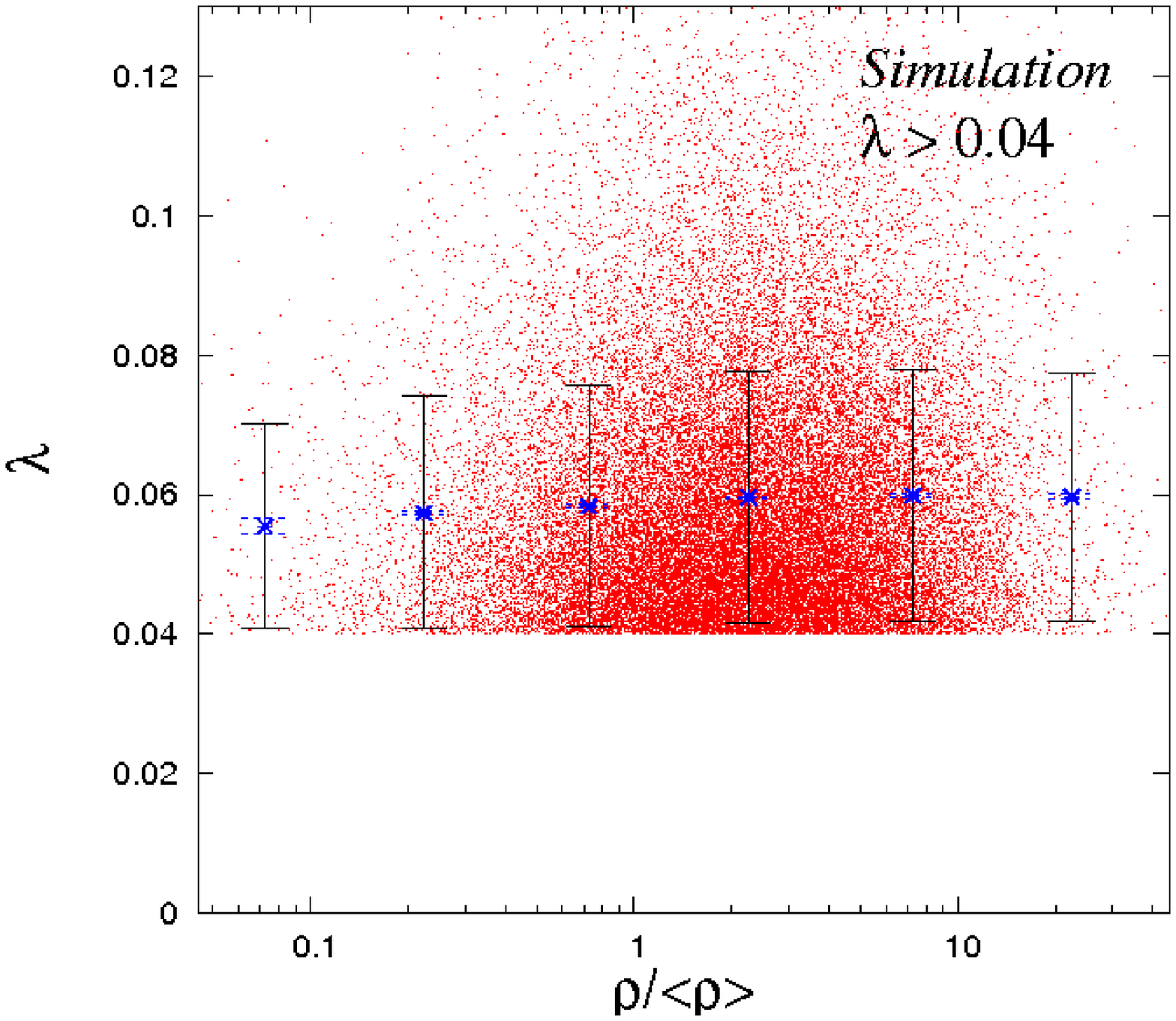}
\end{tabular}
\label{Density}
\caption[ ]{Relations between spin and environmental density, showing the mean $\lambda$ value for each bin as a function of 
the normalized density, the solid error bars corresponds to the dispersion and the broken error bars to the standard error in the calculation of the mean value. Top panels corresponding to the SDSS sample, 
on the left-hand side the entire sample using the 11 597 galaxies and on the right-hand side using only the 7 754 disc galaxies. 
Bottom panels corresponding to the simulated dark matter haloes: on the left-hand side the whole 100 000 sample, on the 
right-hand side 52 950 haloes with $\lambda > 0.04$ }
\end{figure*}

\subsection{Correlations between $\lambda$ and environment density}

To study if there is any correlation between spin and the environment, we use an estimation of the local background density defined 
directly from observations, that is continuous and able to characterize the full range of galaxy environments. It measures 
the local number density of galaxies at a given smoothing scale. 

Being a continuous measure, it does not make any difference
on field, voids, filaments or clusters but their estimation can be done based on the present density measure. The measurement
of the environmental density was done by Park et al. (2007), where they adopt a Spline kernel with 
adaptive smoothing scale to include a fixed number of galaxies within the centrally weighted kernel. The details of the
method using the spline kernel are summarized in Park et al. (2007). The local density at a given location for each 
galaxy is measured by:

\begin{equation}
\label{rho}
\rho(\textbf{x})/<\rho> =  \sum_{i=1}^{20} W_{i}(|\textbf{x}_{i}-\textbf{x}|)/<\rho>,
\end{equation}

using the spline kernel weight $W(r)$ for the background density estimation as in Park, Gott \& Choi (2008).
The mean mass density of the sample is obtained by

\begin{equation}
\label{meanrho}
<\rho> =  \sum_{all} 1/V,
\end{equation}

where the summation is over all galaxies in the sample contained in the total volume $V$.

With this measure for the local density of the environment, we are able to search for any correlation among environment 
and spin. In several studies, the samples of simulated galaxies are divided according to their location in field, voids, 
filaments or clusters, by cuts in the overdensity of the environment or simply by cuts of the normalized local density, 
and then mean values of $\lambda$ for each subsample are obtained or $\lambda_{0}$ values from fits of lognormal functions 
to the distributions at each cut. We divided our sample into six cuts according to the normalized environmental density, and for each cut 
we obtained the mean $\lambda$ value and its dispersion. The result is presented in the upper left-hand panel of 
Fig. 3, where the value 
of total infered $\lambda$ versus $\rho/<\rho>$ for each galaxy is plotted and we show the mean $\lambda$ value for each cut as a function of 
the normalized density, the solid error bars corresponds to the dispersion and the broken error bars to the standard error in the calculation of the mean value, this convention will be kept in the following figures.
As can be seen, there is no clear trend, taking into account the dispersion in each bin.

\begin{figure}
\centering
\begin{tabular}{c}
\includegraphics[width=0.475\textwidth]{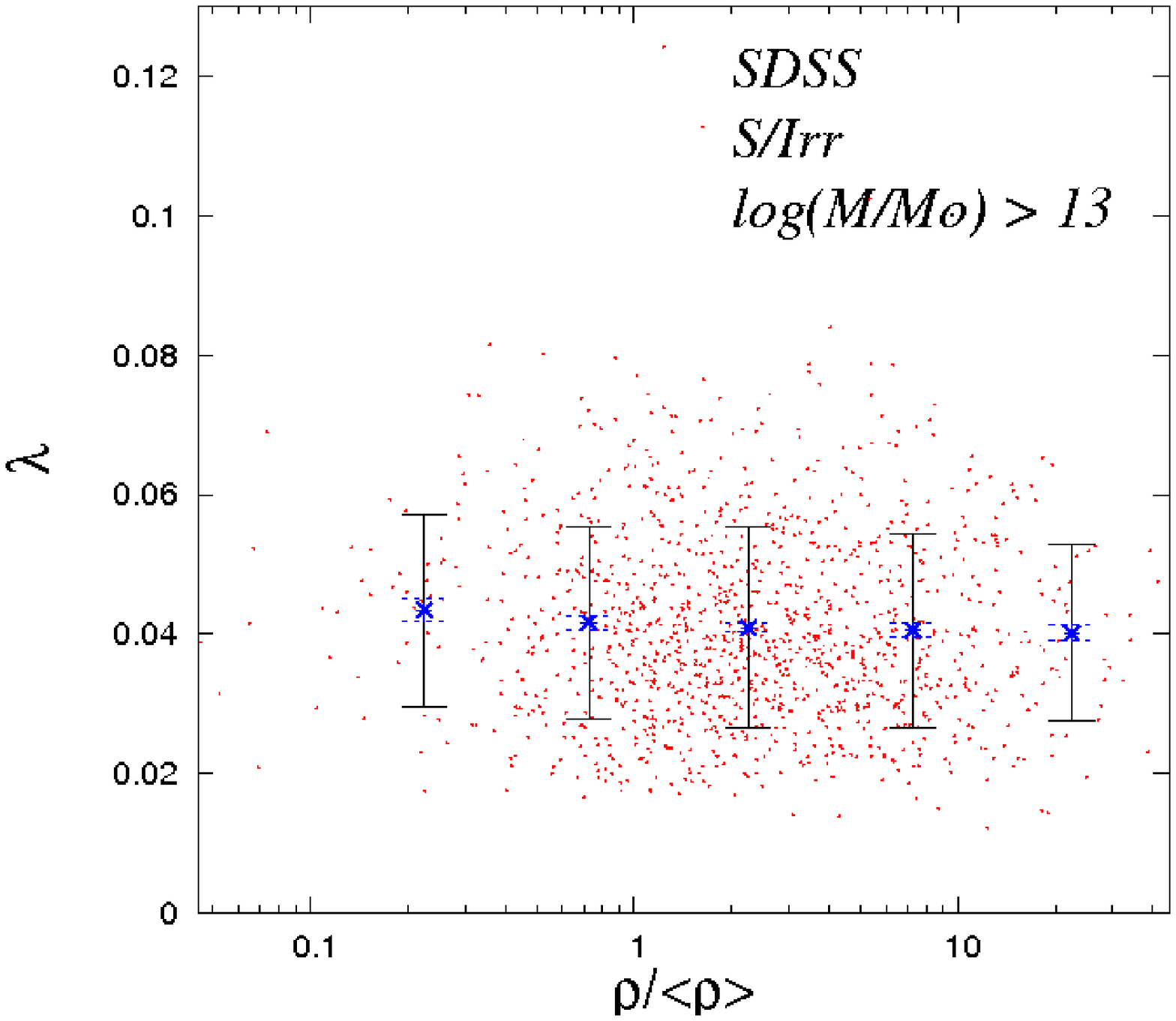}\\
\includegraphics[width=0.475\textwidth]{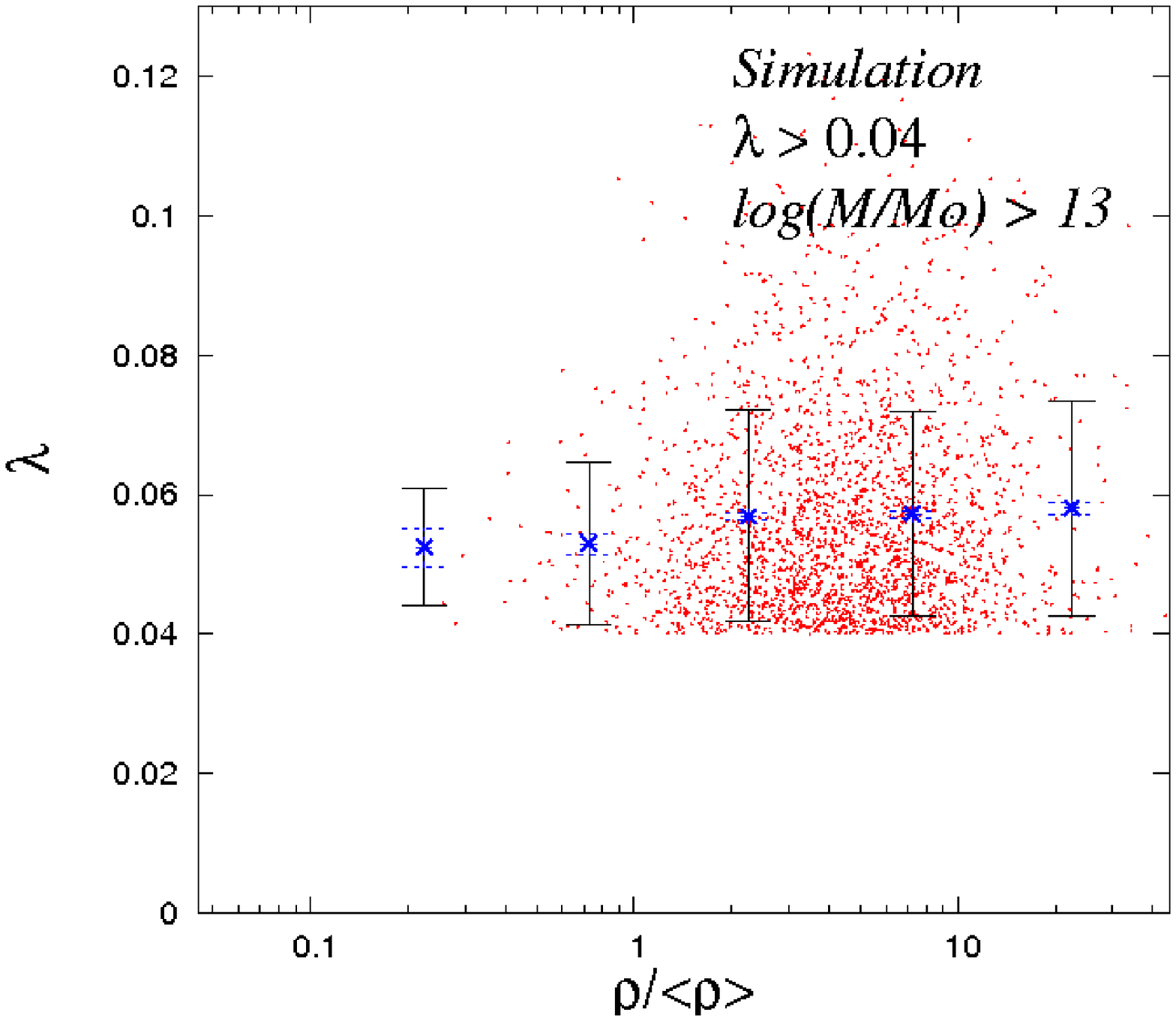}
\end{tabular}
\label{densityHighMass}
\caption[ ]{Relation between spin and the environmental density for massive disc galaxies. Top: 1 238 disc galaxies 
form SDSS sample. Bottom: 3 536 haloes with high $\lambda$ from the N-body simulation, with halo masses in the same range as
the SDSS galaxies shown above.}
\end{figure}

As discussed in Paper II, our $\lambda$ estimation for early-type galaxies is not as accurate as the estimation for disc 
galaxies due to the uncertainties in the way dissipation, pressure support and projection effects are  
taken into account to obtain equation ~\ref{LamObsEll}; with this in mind, we also checked if being more rigorous in 
the estimation of $\lambda$, we could find any trend with the environment. Using only the 7 753 disc galaxies, and 
dividing the sample again in six bins, we obtain the result presented in the upper right-hand panel of Fig. 3, again showing little trend
if any, between $\lambda$ and the normalized local density of the environment.

Using the full sample of haloes from our N-body simulation, divided into six bins and calculating the mean $\lambda$  
value and dispersion for each bin, we obtain the result shown in Fig. 3, bottom left-hand panel. The small change in the mean value 
and large dispersion are consistent with there being no correlation between the two properties, in agreement with what was obtained 
using the SDSS sample. To compare the results arising from the numerical simulation with our sample of disc galaxies, where 
we are more confident of the $\lambda$ estimation, we can take as a first-order criterion for haloes harbouring "late-type galaxies" only the haloes with $\lambda > \lambda_{0}$, where $\lambda_{0}$ comes from the fit of the whole sample to a lognormal function, 
in this case, $\lambda_{0} = 0.0354$. The behavior of the spin for the high spin haloes of the N-body simulation, as a 
function of density, is shown in Fig. 3 bottom right-hand panel, with the same qualitative result as obtained 
using the whole sample. 

Recent works studying the variation of $\lambda$ in different environments, sometimes find a weak dependence for high-mass galaxies,
in the sense that higher spin galaxies are more strongly clustered than lower spin galaxies (Bett et al. 2007; Hahn et al. 2007). 
We inspect if keeping only the high-mass galaxies, we can find such a significant correlation. For both samples, the SDSS and 
the simulation, we took into account disc galaxies with total halo masses $M > 1 \times 10^{13}M_{\sun}$, and simulates haloes in the same mass range. To obtain the 
mass of the SDSS galactic haloes we
used a  mass-luminosity relation which relates the baryonic mass of the galaxy with its absolute magnitude (e.g. Kassin, de Jong \& Weiner 2006),
and we hold the same baryonic to dark matter fraction introduced in Section 2. The behavior found is presented in Fig. 4, 
top panel corresponding to the SDSS sample and bottom panel corresponding to the N-body simulation, where we can see that the 
trend, almost imperceptible in Fig. 3 is now a little more clear but still not significant, taking into 
account the large dispersion in each bin. In the case of the SDSS sample, we find the same behaviour as in Fig. 
3.

To summarize the results of this subsection, we can conclude that we could not find any clear correlation between the 
$\lambda$ spin parameter and the environment where the galaxies are located, as pointed out by several theoretical studies for dark matter haloes in simulations 
(Lemson \& Kauffmann 1999, Maccio et al. 2007, Bett et al. 2007). Neither at low nor high masses, not for real galaxies or simulated haloes,
do we find any significant trend in the parameters describing the distribution of $\lambda$ with environment density. Up to
this point, current cosmological modeling appears fully consistent with real galaxies.

\subsection{Correlations between $\lambda$ and halo mass}

\begin{figure*}
\centering
\begin{tabular}{cc}
\includegraphics[width=0.475\textwidth]{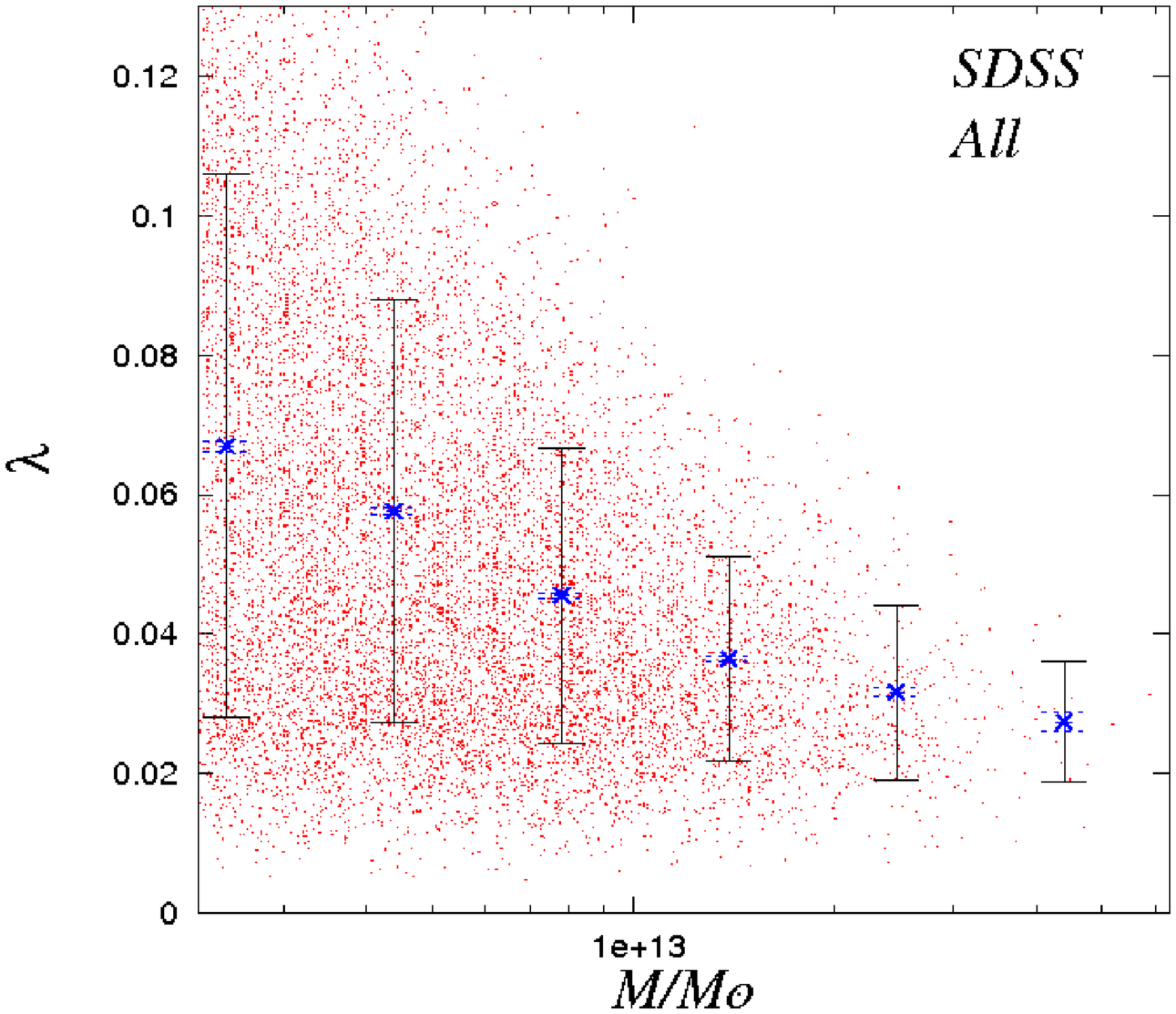} & \includegraphics[width=0.475\textwidth]{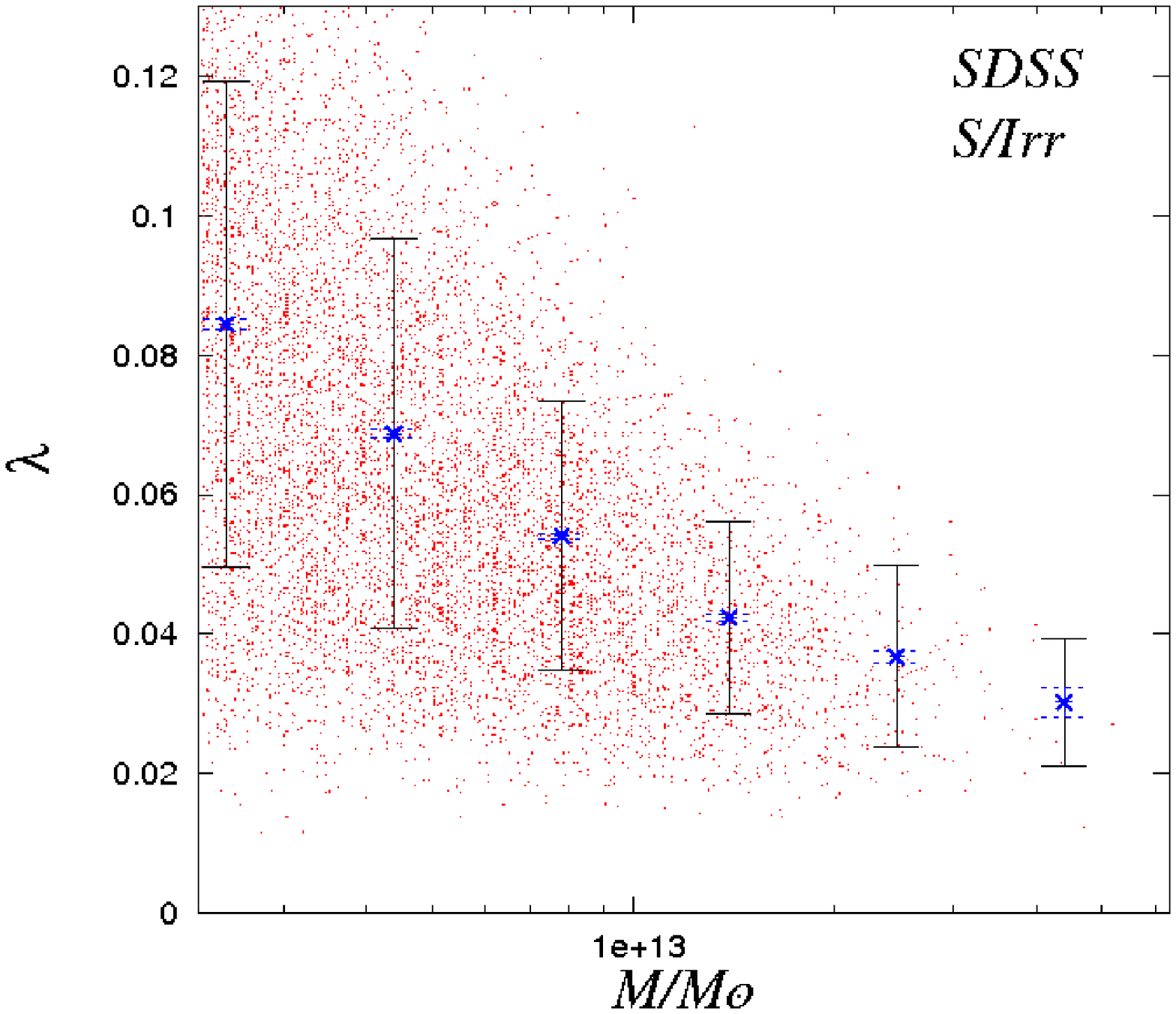} \\
\includegraphics[width=0.475\textwidth]{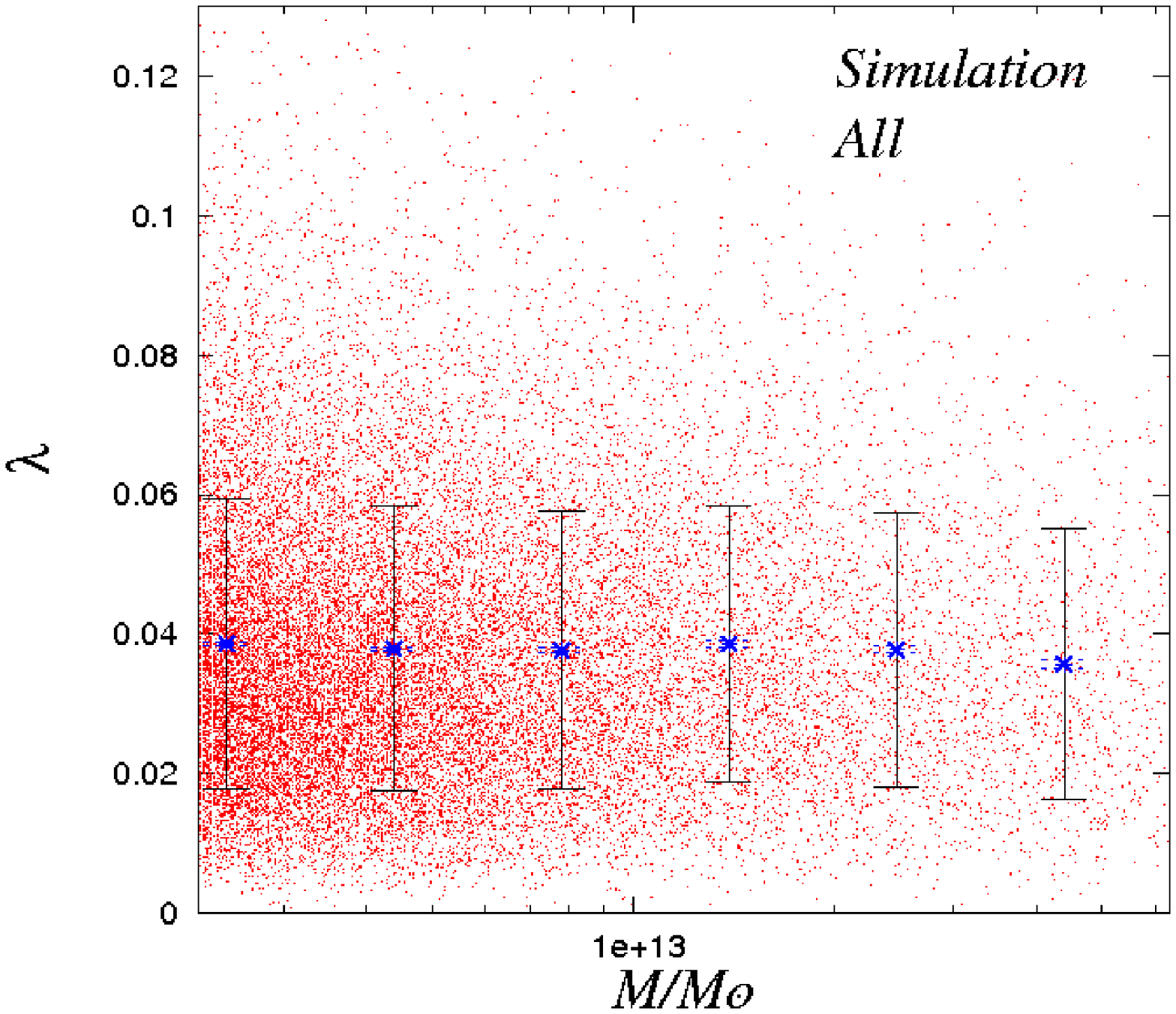} & \includegraphics[width=0.475\textwidth]{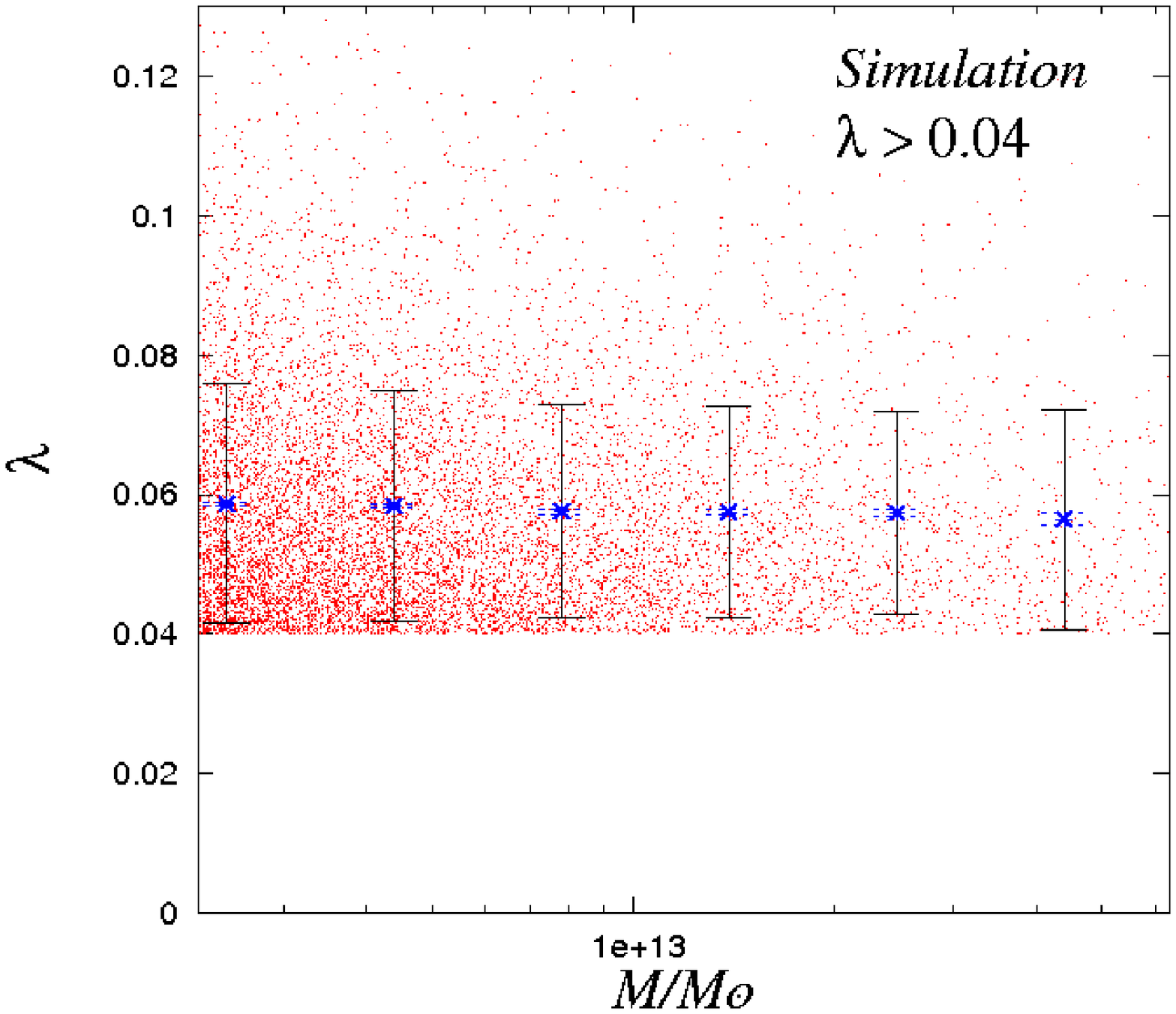}
\end{tabular}
\label{fig:Mass}
\caption[ ]{Relations between spin and mass. Top panels corresponding to the SDSS sample: on the left-hand side the entire 
sample using the 11 597 galaxies and on the right-hand side using only the 7 754 disc galaxies. Bottom panels corresponding 
to the simulated dark matter haloes: on the left-hand side the whole 100 000 galaxies, on the right-hand side 52 950 galaxies with $\lambda > 0.04$ }
\end{figure*}

After having found no correlation between $\lambda$ and the local environment, we now search 
for a correlation with internal properties of the galaxies. In many analytic and numerical studies, $\lambda$ has been established 
as an important, if not the most important, parameter determining the galactic type (Fall \& Efstathiou 1980; Flores et al. 1993;
Firmani, Hernandez \& Gallagher 1996; Silk 2001). In that sense, we expect it to correlate with the most important structural 
parameters such as thickness of the disk, bulge-to-disc ratio, colours, metallicity among others. van den Bosch (1998) emphasizes 
the important role of this parameter in explaining the origin of the Hubble sequence, and shows correlations of $\lambda$ with the
bulge-to-disc ratio, in the sense that systems with high $\lambda$ presents systematically lower bulge-to-disc ratios. 
Kregel et al. (2005) explain the role of $\lambda$ in the stellar velocity dispersion and predict extended thin
disc in galaxies immersed in high spin angular momentum dark matter haloes. Trends with star formation efficiencies, gas 
fractions, metallicities, abundance gradients and colours have also been predicted in several studies 
(e.g. Dalcanton et al. 1997; Boissier et al. 2001; Churches, Nelson \& Edmunds 2001). In Paper I and in Cervantes-Sodi \& Hernandez (2008), we showed this
type of scaling but using samples of real galaxies, confirming the general results of theoretical studies.

But still, more important than all these properties is the mass of the galaxy. The results arising from early works are rich 
and varied, some of them pointing in the sense that no correlation is found between $\lambda$ and mass (e.g. Warren et al. 1992; 
Lemson \& Kauffmann 1999; Shaw et al. 2006; Maccio et al. 2007) and others showing from weakly to marked trends between these 
two parameters, again, always for cosmological simulations (e.g. Cole \& Lacey 1996; Jang-Codell \& Hernquist 2001 - although the low resolution of this particular study might make this results questionable, Bett et al. 2007).
We showed in Papers I and II a strong correlation between halo $\lambda$ as estimated through equation (\ref{LamObs}), and
visually assigned morphological type, in the expected sense, with later type galaxies corresponding to high values of inferred $\lambda$. 
Given the tight correlation that exists between Hubble type and mass, where high mass systems tend to be of earlier types,
it is natural to expect a trend to appear when cutting the sample of $\lambda$ as a function of mass. We expect high-mass
galaxies to be found preferentially amongst the low-$\lambda$ haloes.

We investigate whether our SDSS sample presented this correlation or not. Once total $\lambda$ values were obtained for all the galaxies using 
the corresponding equation, and using the TF relation presented in the preceding subsection, we obtained the upper 
left-hand panel of Fig. 5,
plotting $\lambda$ as a function of $mass$. In this case, the trend is very clear, low-mass galaxies present typically 
higher $\lambda$ values and larger dispersion, exactly what the association of $\lambda$ and Hubble type mentioned previously
would have suggested. We note that precisely such a trend is reported in the theoretical study of Maller et al. (2002) for
the case of a tidal-torque scenario for the acquisition of angular momentum, as opposed to the scenario where
angular momentum grows through the merger of satellites. We take our results as evidence of a mechanism of acquisition
of angular momentum for galaxies, where it is the ambient tidal field what torques up a halo, with little participation
of repeated mergers.

The lower left-hand panel in Fig. 5 shows halo $\lambda$ as a function of $mass$ for the simulation, 
although a certain trend is present in the same sense as in the SDSS sample, its magnitude is small and the dispersion large.

Our estimates of $\lambda$ are more precise for disc galaxies, in both parameters, the spin and the mass. Taking only 
this galaxies, the corresponding figures for the SDSS sample and the above average $\lambda$ simulated haloes are presented in Fig. 5, 
right-hand panels. In the case of the SDSS sample, the trend found using the hole sample, disc galaxies plus ellipticals, 
appears even stronger. The response in the case of the simulation is similar, the trend is kept but the dispersion 
decreases for larger galaxies, as in the SDSS sample but at a much lower level.

\section{Discussion}

We have presented our results in the study of a large sample of galaxies taken from the SDSS and compared to the situation found in
recent cosmological N-body simulations. We looked for correlations of the total galactic $\lambda$ spin parameter distribution
with both halo mass and local environmental density. With the local environment density we find 
no correlation, as reported in several theoretical studies and reproduced here using a cosmological N-body simulation. The large 
dispersion and little trend seen when dividing the samples into bins with different normalized environment density, are 
consistent with no correlation at all for a $\lambda - <\rho>$ relation, for both; infered haloes of observed galaxies and modelled haloes.

The case of the $\lambda - mass$ relation is quite different. Analyzing our SDSS sample, we notice that the mean value 
of $\lambda$ tends to decrease as the mass increases. But not only the mean value, also the dispersion of $\lambda$ tends to
 decrease as the value of the mass increases. The effect is stronger using only the disc galaxies, where our estimate 
is more precise. Large galaxies are seen to form a more coherent low-$\lambda$ sample with small dispersion, while
small galaxies show mean values of $\lambda$ of about a factor of 3 larger than what is found for our most massive galaxies. 
The dispersion about mean values shows the same trend, with small galaxies in our SDSS sample having a much larger
dispersion than the large galaxies. Analyzing the N-body simulation we found a very weak trend in the same 
sense using the high spin galaxies, which would correspond to disc galaxies. 

Burkert (2003), found a correlation between the spin parameter and the baryon fraction of the form $\lambda \propto F^{2/3}$. 
In our simple model used to calculate $\lambda$ from observable parameters, the dependence is even stronger; $\lambda \propto F$,
this could be invoked to explain the relation found between mass and $\lambda$, strong star formation can drive galactic winds 
which clear a fraction of the baryonic material, preferentially in small systems. This would lead to 
a lower effective baryon fraction $F$ in smaller systems, and hence, the assumption of a constant baryon fraction would lead 
us to overestimate $\lambda$ in small systems. However, as has been shown recently
by several authors studying the accretion and expulsion of baryons in cosmological scenarios (e.g. Crain et al. 2007; Brooks 2007), 
the galactic baryon fraction is almost constant in the range of masses we are dealing with in this work. A fractional
variation in $F$ of over a factor of 3, over the 1.5 orders of magnitude in galactic mass we are dealing with here, is inconsistent with the physics of galaxy formation. We excluded galaxies of smaller masses,
partly to guarantee completeness of the sample, and partly to stay well above the limit where galactic winds can be expected
the change the baryon fraction differentially.

If the baryonic component is susceptible to angular momentum dissipation or baryonic to dark halo angular momentum transfer, 
and this effect is sharper in more massive systems, the trend observed 
with the mass could be due to a subestimation of the halo spin. Again, we can discard this hypothesis based on the study of 
Kaufmann et al. (2007), in which the loss of angular momentum of disc particles on a smoothed particle hydrodynamic simulation 
is traced. They found that almost half of the initial angular momentum is lost via different mechanisms in low-resolution 
cosmological simulations, but when a high-resolution simulation is used only 10-20 per cent of the original angular momenta 
is lost, not enough to explain the strong correlation presented here between mass and $\lambda$. Further, no trend with mass
is seen.

van den Bosch et al. (2002), using a numerical simulation of structure formation in a $\Lambda CDM$ cosmology, concluded that 
the detailed angular momentum distributions of the gas and dark matter components in individual haloes are remarkably similar, 
and the differences between them present no significant dependence on the halo virial mass, result that supports the model 
used to calculate $\lambda$. A similar conclusion
was obtained by Zavala, Okamoto \& Frenk (2008). Tracing the evolution of the specific angular momenta of the dark matter and 
baryonic components on a cosmologically forming disc galaxy, they show that the baryonic component tracks the 
specific angular momentum of the halo, 
and their actual values are very similar throughout the entire history of the galaxy. The lack of a strong trend with mass
in the effect of baryonic angular momentum transfer or dissipation found in detailed hydrodynamical simulations, shows that the
correlation we find in mean values of $\lambda$ which decrease with increasing galactic halo mass, is real. 
In addition, the relation between $R_{d}$, $V_{d}$ and $\lambda$ of equation ~\ref{LamObs} is practically the same as what is found in the results of the high-resolution simulation of Okamoto et al. (2005), as seen in Fig. 1.

To summarize, we find a good accordance between the integral distributions of $\lambda$ inferred from the large SDSS
sample we studied, and results from current cosmological simulations, both in terms of the functional form of the distributions,
and the parameters describing these distributions. The same is the case when comparing the samples as a function of
environment density, with no clear trend apparent. However, when looking at the samples as a function of mass, no clear
trend is seen in the simulated haloes, but in the sample of real galaxies we find a strong tendency for both mean values and dispersions in
$\lambda$ to decrease with increasing galactic mass; Tonini et al. (2006) reached similar conclusions through a more indirect study of galactic populations, and in order to explain the infered baryonic scaling relations for a sample of around 80 isolated disc galaxies, using models of galactic evolution, Avila-Reese et al. (in preparation) required an anticorrelation between the mass, and both, the spin and the baryon fraction. This provides constrains on the form of acquisition of the angular momenta 
in galactic systems and important clues to discriminate cosmological models, given the dependency of this relations on the 
cosmology assumed, e.g. Eke, Efstathiou \& Wright (2000). Comparing with existing studies of galaxy formation, we see that the trend 
we find cannot be ascribed to astrophysical effects associated with baryonic loss or dissipation. It must rather be 
understood as evidence in favor of an angular momentum acquisition mechanism where ambient tidal fields torque up galactic 
haloes as a whole at high redshift, rather than the progressive spin up caused by a repeated and prolonged satellite accretion process.
To first order, the trend for a lower mean value of $\lambda$ at larger halo masses can be understood by thinking of a constant
ambient tidal field, which then results in higher $\lambda$ for smaller systems.

\section*{Acknowledgments}

We thank the anonymous referee for pointing out a number of impressions present in the original version of the manuscript. The work of B. Cervantes-Sodi was supported by a CONACYT scholarship. The work of XH was partially supported by 
DGAPA-UNAM grant no IN114107. We would like to thank J. Koda for providing us the luminosity profiles of his simulated 
galaxies, to perform the test of our model. We would like to thank Dr. Yun-Young Choi for preparing the SDSS data for us.
CP acknowledges the support of the Korea Science and Engineering
Foundation (KOSEF) through the Astrophysical Research Center for the
Structure and Evolution of the Cosmos (ARCSEC).

Funding for the SDSS and SDSS-II has been provided by the Alfred P. Sloan
Foundation, the Participating Institutions, the National Science
Foundation, the U.S. Department of Energy, the National Aeronautics and
Space Administration, the Japanese Monbukagakusho, the Max Planck
Society, and the Higher Education Funding Council for England.
The SDSS Web Site is http://www.sdss.org/.

The SDSS is managed by the Astrophysical Research Consortium for the
Participating Institutions. The Participating Institutions are the
American Museum of Natural History, Astrophysical Institute Potsdam,
University of Basel, Cambridge University, Case Western Reserve University,
University of Chicago, Drexel University, Fermilab, the Institute for
Advanced Study, the Japan Participation Group, Johns Hopkins University,
the Joint Institute for Nuclear Astrophysics, the Kavli Institute for
Particle Astrophysics and Cosmology, the Korean Scientist Group, the
Chinese Academy of Sciences (LAMOST), Los Alamos National Laboratory,
the Max-Planck-Institute for Astronomy (MPIA), the Max-Planck-Institute
for Astrophysics (MPA), New Mexico State University, Ohio State University,
University of Pittsburgh, University of Portsmouth, Princeton University,
the United States Naval Observatory, and the University of Washington.

\end{document}